\documentclass[aps, prd, amsmath, floats, floatfix, twocolumn,superscriptaddress, nofootinbib, showpacs, letterpaper, preprintnumbers]{revtex4-1}
\usepackage{graphicx}
\usepackage{amsmath,amssymb}
\usepackage{amsfonts}
\usepackage{xspace} 
\usepackage[usenames]{color}
\usepackage{dcolumn}
\usepackage{bm}

\usepackage{ulem}
\newcommand{\Cornell}{\affiliation{Center for Radiophysics and Space
    Research, Cornell University, Ithaca, New York, 14853, USA}}

\newcommand{\Caltech}{\affiliation{Theoretical Astrophysics 350-17,
    California Institute of Technology, Pasadena, CA 91125, USA}}

\definecolor{darkgreen}{rgb}{0.2,0.7,0.2}

\begin{document}

\title{Simulating merging binary black holes with nearly extremal spins}

\author{Geoffrey Lovelace}\Cornell
\author{Mark A. Scheel}\Caltech
\author{B\' ela Szil\' agyi}\Caltech

\begin{abstract}
Astrophysically realistic black holes may have spins that are nearly extremal 
(i.e., close to 1 in dimensionless units). 
Numerical simulations of 
binary black 
holes are important tools both for calibrating analytical 
templates for gravitational-wave detection and for exploring the 
nonlinear dynamics of curved spacetime.
However, all previous simulations of binary-black-hole inspiral, 
merger, and ringdown have been
limited by an apparently insurmountable barrier: the merging holes' spins
could not exceed $0.93$, which is still a long way from the maximum possible
value in terms of the physical effects of the spin.
In this paper,
we surpass this limit for the first time, opening the way to explore 
numerically the behavior of merging, nearly extremal black holes.
Specifically, using an improved initial-data method suitable for 
binary black holes with nearly extremal spins,
we simulate the inspiral 
(through 12.5 orbits), merger and ringdown of two 
equal-mass 
black holes with equal spins of magnitude 0.95 antialigned with the orbital 
angular momentum. 
\end{abstract}

\date{\today}

\pacs{04.25.dg, 04.30.-w}

\maketitle

\section{Introduction}
\label{sec:Introduction}

Although there is considerable uncertainty, it is possible that 
astrophysical black holes exist with 
nearly extremal spins (i.e., in dimensionless units spins close to 1,
the theoretical upper limit for a stationary black hole). 
Binary black hole (BBH) mergers 
in vacuum typically lead to 
remnant holes with dimensionless spins 
$\chi\sim 0.7-0.8$~\cite{Tichy2008,Sperhake2008Ecc,Lousto2009},
although if the 
merging holes 
are surrounded by matter 
the remnant's spin typically 
could be 
higher than $\chi\sim 0.9$~\cite{Tichy2008,Lousto2009}.
Black holes can reach higher spins via 
prolonged
accretion~\cite{VolonteriEtAl:2005,BertiVolonteri:2008}:
thin accretion disks (with magnetohydrodynamic effects neglected) lead 
to spins as large as $\chi \sim 0.998$~\cite{Thorne:1974}, 
while thick-disk 
accretion with magnetohydrodynamic effects included can lead to spins 
as large as 
$\chi\sim 0.95$~\cite{GammieEtAl:2004,Shapiro:2005}. 
Even without accretion, at very high mass ratios with spins aligned with the 
orbital 
angular momentum, binary black hole mergers can also lead to holes
with nearly extremal spins~\cite{Rezzolla:2007xa,Kesden2008,KesdenEtAl:2010}.
There is observational evidence suggesting the 
existence of black holes with 
nearly extremal spins in quasars~\cite{WangEtAl:2006}, and 
some efforts to infer the spin of 
the black hole in microquasar GRS 1915+105 from its x-ray spectra suggest 
a spin larger than 0.98, though other analyses suggest the 
spin may be much lower~\cite{McClintockEtAl:2006,Middleton:2006,Blum:2009}.

\begin{figure}[t]
\includegraphics[width=3.5in]{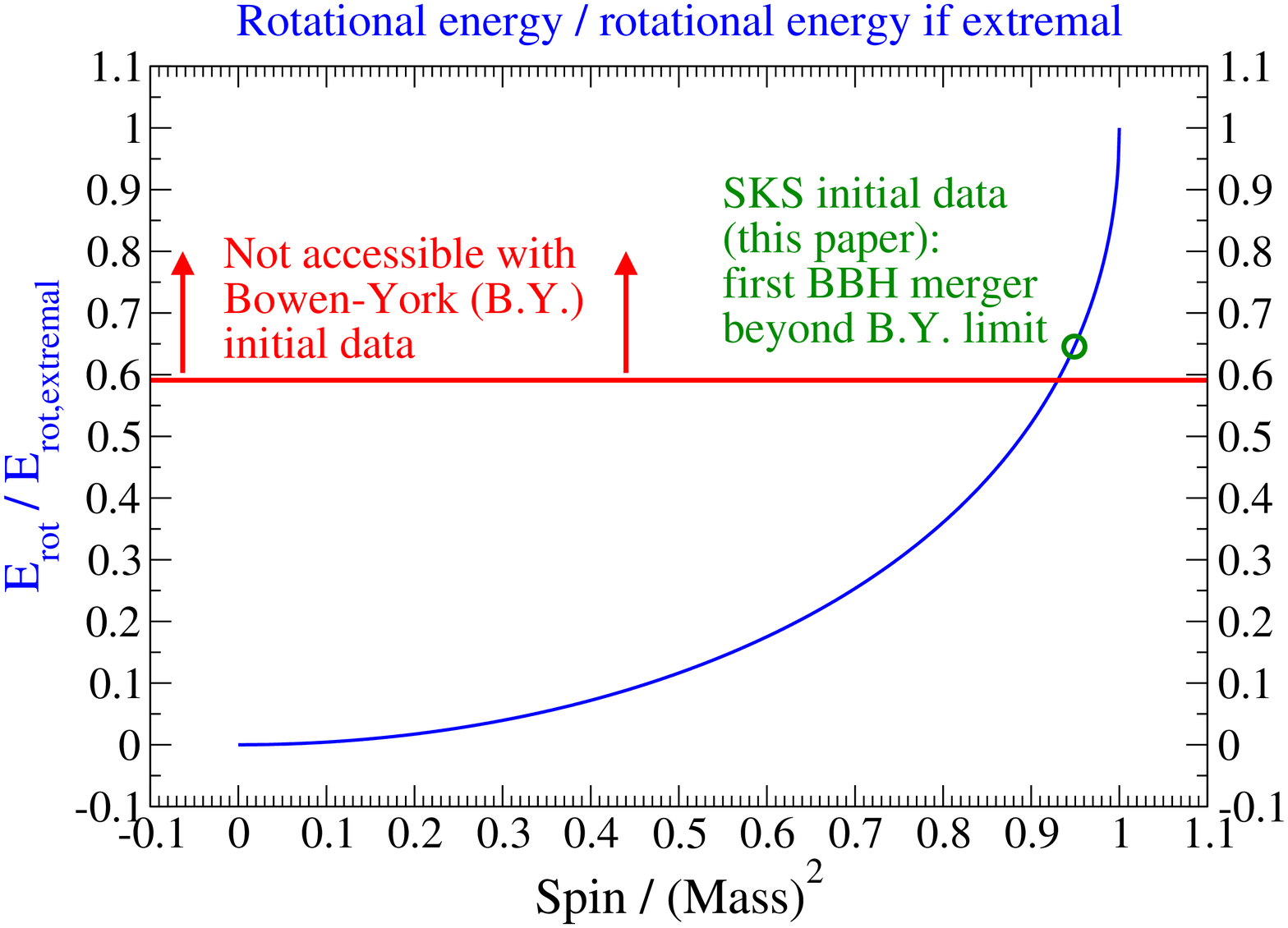}
\caption{The rotational energy of a Kerr black hole as a function of the 
hole's dimensionless spin parameter $\chi:=\mbox{Spin}/(\mbox{Mass})^2$. 
The thick red line indicates 
the Bowen-York limit: standard Bowen-York 
puncture initial data---used in almost all numerical binary-black-hole  
calculations to date---cannot yield rotational energies more than 
60\% of the way to extremality.
By using instead initial data based on two superposed Kerr-Schild holes 
(``SKS initial data''), in this paper we surpass the Bowen-York limit 
(green circle), opening the way 
for numerical studies of merging, nearly extremal black holes.
\label{fig:RotEnergy}}
\end{figure}

Merging BBHs---possibly with nearly extremal spins---are 
among the most 
promising sources of gravitational waves for current and future 
detectors. Numerical simulations of BBHs are important tools both for 
predicting the gravitational waves that detectors will observe
and for exploring 
the behavior of nonlinear, highly dynamical spacetimes. Following 
Pretorius' breakthrough in 2005~\cite{Pretorius2005a}, several 
groups have successfully simulated the inspiral, merger, and ringdown of 
two coalescing 
black holes in a variety of initial configurations; however, 
all prior BBH simulations have
been limited to spins of 0.93 or less, which is quite far from 
extremal. The parameter $\chi$ is a poor measure of how close a black hole is to extremality in terms of physical effects:
a black hole with spin 0.93 has less than 60\% of the rotational 
energy of an extremal hole with the same mass (Fig.~\ref{fig:RotEnergy}).

Previous simulations have been 
unable to reach higher spins because of the way they construct their 
initial data.
Just as initial data for Maxwell's equations must satisfy 
constraints (the electric and 
magnetic fields must have vanishing divergence in vacuum), 
initial data 
for the Einstein equations must satisfy constraint equations.
Most BBH simulations begin with 
puncture initial data~\cite{Brandt1997}, which assumes that the 
initial spatial metric is \emph{conformally flat} (i.e. 
proportional to the metric of flat space).
With this assumption, 3 of the 4 constraint equations can be solved 
analytically using the solutions of 
Bowen and York~\cite{bowen79,Bowen-York:1980}; however, 
conformally flat
initial data cannot describe single or binary black holes that are both 
in equilibrium and possess linear~\cite{york80} or 
angular~\cite{GaratPrice:2000,Kroon:2004} momentum.
Bowen-York puncture data \emph{can} yield solutions of binary black holes 
with spins as large as $\chi=0.984$ initially, but when such initial data are 
evolved, 
the holes quickly relax to spins of 
about $\chi=0.93$ 
or less~\cite{cook90,DainEtAl:2002,HannamEtAl:2009}.
Several groups have evolved BBH puncture data with 
spins near but below this \emph{Bowen-York 
limit}~\cite{Rezzolla:2007xa,HannamEtAl:2010,MarronettiEtal:2008},
with the simulation by 
Dain, Lousto, and Zlochower~\cite{DainEtAl:2008} coming the closest 
with spins of $0.967$ at time $t=0$ quickly falling to 
0.924.

To reach spins beyond the Bowen-York limit,
one must 
begin with initial data that is \emph{conformally curved}. 
Recently, 
Liu and collaborators~\cite{Liu:2009} have constructed and evolved
conformally curved initial data based on that of 
Brandt and Seidel~\cite{BrandtSeidel:1995a,BrandtSeidel:1996} 
for a single black hole with spins as high as 
$\chi=0.99$.
Hannam and collaborators~\cite{Hannam2007b} have constructed and evolved 
conformally curved BBH initial data~\cite{Dain:2001a,Dain:2001b} 
for head-on mergers of black holes with spins as large as $\chi=0.9$. 
In Ref.~\cite{Lovelace2008}, conformally curved BBH data with  
spins of $\chi=0.93$ were constructed and evolved through the 
first 1.9 orbits of an inspiral, but no attempt was made to simulate the 
complete inspiral, merger, and ringdown.

In this paper, we 
demonstrate that conformally curved initial data 
is suitable for simulations with 
nearly extremal spins by using it to compute the first
inspiral, 
merger, and ringdown of two black holes 
with 
spins larger than the Bowen-York limit.
By surpassing this
limit, our results open the way for numerical 
exploration of the gravitational waveforms and nonlinear dynamics 
of black holes that are nearly extremal.

\label{sec:InitialDataMethods}
\begin{table}
\begin{tabular}{|lr|lr|lr|}
\hline
$M_i/M$ & 0.5000
& $M_{\rm ADM}/M$ & 0.9933 &
$d_0/M$ & 15.366\\ 
$\chi^z_i$ & -0.9498 & $J^z_{\rm ADM}/M^2$ & 0.6845
& $\Omega_0 M$ &  0.014508 \\
& & & &  $\dot{a}_0$ & -0.0007139 \\
\hline
\end{tabular}
\caption{Properties of initial data evolved in this paper.
The quantity $M$ denotes 
the sum of the holes' Christodoulou masses at $t=0$.
Hole $i$ (where $i=A\mbox{ or }B$) has Christodoulou mass $M_i$ and 
dimensionless spin $\chi^z_{i}$ along the 
z axis (i.e., in the direction of the orbital angular momentum). 
Also listed is the 
Arnowitt-Deser-Misner (ADM) mass $M_{\rm ADM}$ and angular momentum 
$J^z_{\rm ADM}$ (e.g., Eqs. (25)--(26) of Ref.~\cite{Lovelace2008}).
The initial angular velocity $\Omega_0$, 
radial velocity $\dot{a}_0$, and coordinate separation 
$d_0$ were tuned to reduce the orbital eccentricity. 
\label{tab:ID}}
\end{table}

\section{Initial data}
We 
evolve a
low-eccentricity
initial data set: 
a BBH where the holes have  
equal masses and equal spins of magnitude 0.95 antialigned with the orbital 
angular momentum. Some properties of the initial data used 
in this paper are listed in Table~\ref{tab:ID}.

Following Ref.~\cite{Lovelace2008} and the references therein, 
we construct constraint-satisfying initial data by solving
the extended 
conformal thin sandwich equations with 
quasiequilibrium boundary conditions~\cite{York1999,Cook2002,Cook2004,
Caudill-etal:2006,Gourgoulhon2001,Grandclement2002} using 
a spectral elliptic solver~\cite{Pfeiffer2003}. 
The initial spatial metric 
is proportional to a weighted superposition of the metrics of two 
boosted, spinning Kerr-Schild black holes. 

We measure the quasilocal spin $S_{\rm AKV}$ of each hole in the initial 
data using the 
approximate-Killing-vector method summarized 
in Appendix A of Ref.~\cite{Lovelace2008}, which is very similar to the 
prescription previously 
published by Cook and Whiting~\cite{Cook2007}. The dimensionless spin 
of each hole $\chi$ is then related to $S_{\rm AKV}$ by the 
formula
$\chi := S_{\rm AKV}/M_{\rm chr}^2$,
where $M_{\rm chr}:=\sqrt{M_{\rm irr}^2 + S^2/4M_{\rm irr}^2}$ 
is 
the Christodoulou mass, $M_{\rm irr}:=\sqrt{A/16\pi}$ is the irreducible 
mass, and $A$ is the area of the horizon. 
(For a single Kerr black hole, $M_{\rm chr}$ reduces to the 
usual Kerr mass parameter.) 

To reduce eccentricity, we follow the 
iterative method of Ref.~\cite{BuonannoEtAl:2010}, 
which is an improvement of the earlier method of 
Ref.~\cite{Pfeiffer-Brown-etal:2007}. 
For each iteration, we construct an initial data set 
and evolve it for 
approximately 3 orbits. Then, the initial 
angular and radial motion of the holes are adjusted to minimize 
oscillations in the orbital frequency. Using this method, we reduce 
the orbital eccentricity to approximately $10^{-3}$. 

\label{sec:EvolutionMethods}
\section{Evolution}
We evolve our 
initial data using 
the Spectral Einstein Code {\tt SpEC}~\cite{SpECwebsite}. 
Building on the 
methods of Ref.~\cite{Szilagyi:2009qz} and the 
references therein, we have 
made several technical improvements to our code which 
both enable us to evolve our $\chi=0.95$ initial data through merger and 
make our code more robust in general.
Here we briefly summarize some of the most 
important improvements;
full details of these techniques will be described in 
a future paper.

We use a computational
domain with the singularities 
inside the horizons excised, and we use
a 
time-dependent coordinate mapping to keep the excision 
boundaries
inside the individual apparent horizons as the horizons orbit and
slowly approach each other~\cite{Scheel2006}.
Our 
coordinate mapping also ensures that the excision surfaces' shapes 
conform to those of the horizons which enclose them. 
One important ingredient of our improved binary-black-hole evolutions is that 
the coordinate
mapping is adjusted {\em adaptively} throughout the 
evolution, which is helpful
because the horizons' dynamics change from 
slow to fast during the simulation.

Because we apply no boundary condition on 
the excision surfaces, 
these surfaces must be pure-outflow boundaries (i.e., must have no 
incoming characteristic fields) in order for the evolution to be well posed.
A second improvement to our code is that we now can
adjust the velocity of each excision surface
to keep the characteristic fields outgoing there.
For the $\chi=0.95$ simulation considered here, 
this characteristic speed control is necessary 
only during the last orbit before merger; earlier, 
it is sufficient to control the size of the excision surface using 
the method of Ref.~\cite{Szilagyi:2009qz}.

A third element which we have recently added to {\tt SpEC}
is spectral adaptive mesh refinement. 
During the evolution, we monitor the truncation error of each 
evolved field, the resolution requirements of the apparent horizons, 
and the local magnitude of constraint violation; to maintain a 
desired accuracy, we then add or remove spectral basis functions as needed.
In the simulation presented in this paper, 
we use spectral adaptive mesh refinement 
only during the final quarter orbit before merger.
Throughout the entire simulation, we also adaptively 
adjust the resolution of the apparent horizon finder as the horizon 
becomes more distorted.

\label{sec:Results}
\begin{figure}
\includegraphics[width=3.2 in]{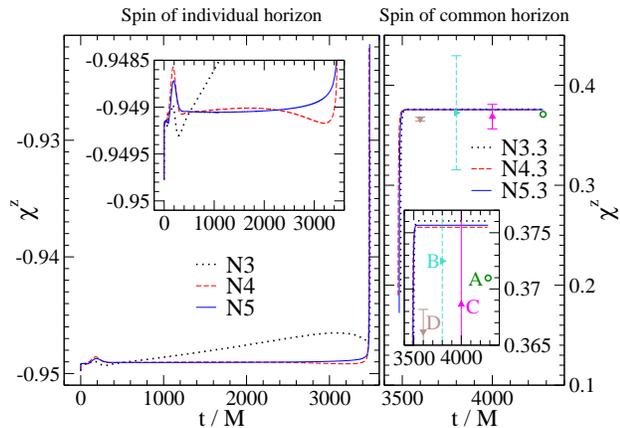}
\caption{\emph{Color online.} 
Left panel: The $z$ component $\chi^z$ 
of the dimensionless quasilocal spin of one individual 
horizon vs time $t$. (The individual holes' spins 
are equal within numerical error.) 
Right panel: spin of the common horizon vs. time and
the final spins predicted by the fitting formulae in
Ref.~\cite{Campanelli2006c} (``A''), Ref.~\cite{Tichy2008} (``B''),  
Ref.~\cite{Barausse2009} (``C''), 
and Ref.~\cite{Rezzolla:2007xa} (``D''),
and (for ``B'',``C'', and ``D'') the 
error bars corresponding to the fitting formulae's listed uncertainties. 
(Note that the horizontal positions of points ``A''--``D'' 
on the figure are arbitrary.)
Our results are shown for several resolutions (labeled $Nx$ or $Nx.y$, 
where $x\in 3,4,5$ and 
$y\in 1,2,3$ label 
the resolution used before and after merger, respectively).
\label{fig:spin}}
\end{figure}
\begin{figure}
\includegraphics[width=3.2 in]{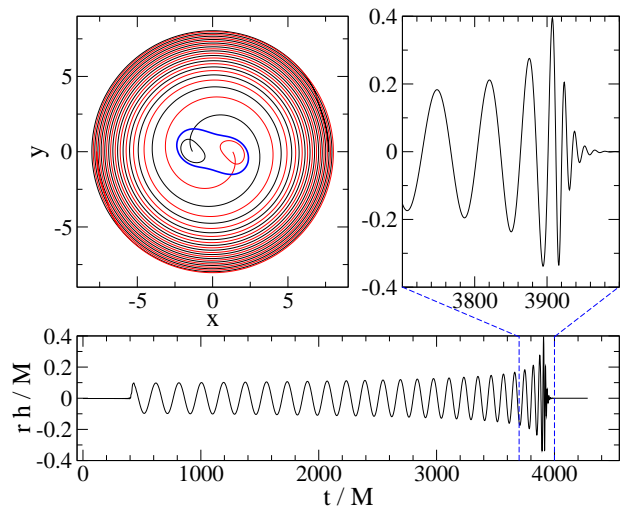}
\caption{\emph{Color online.} The orbital trajectory of the centers of the 
individual horizons and the individual and common horizons 
at the end of the inspiral 
(top left) and the real part of the $\ell=2,m=2$ mode 
of the emitted gravitational waveform $h$ extracted at radius 
$r=405 M$ (bottom). The holes travel through about 12.5 orbits before merging. 
All data is from resolution $N5.3$.
\label{fig:orbits}}
\end{figure}
\section{Results}
In Fig.~\ref{fig:spin}, 
we plot the dimensionless quasilocal spin $\chi$ 
measured on one individual horizon and also on the common horizon. 
From 
$t=0$ to $t=50 M$, there is a sharp, 
numerically resolved drop 
in the magnitude of the dimensionless spin $\chi$ from 0.9498 to 0.9492. 
During the remainder of the inspiral, the spin drifts, with 
the amount of drift decreasing as resolution increases; at the 
highest resolution ($N5$), the spin remains 
$\chi=0.949$ throughout the next 11.8 orbits, 
until just before merger, when
the magnitude of the spin of each hole drops sharply. 
This result demonstrates that it is now possible 
to simulate BBHs where 
the holes retain spins beyond the Bowen-York limit 
throughout the inspiral; it also opens the way for future explorations 
of the strong-field dynamics of merging, nearly extremal holes---dynamics 
that can only be explored using numerical simulations.

During the ringdown, the spin $\chi$ of the common horizon 
quickly relaxes to its final value
of $\chi=0.3757\pm0.0002$ (where the uncertainty is estimated as 
the difference 
between the highest and second-highest resolutions). 
This is approximately consistent 
with but slightly larger than the predictions obtained by 
extrapolating fitting formulae from simulations with lower initial spins of 
Ref.~\cite{Campanelli2006c} ($\chi_{\rm fit}\approx 0.371$), 
Ref.~\cite{Tichy2008} ($\chi_{\rm fit} = 0.372\pm 0.057$),
Ref.~\cite{Barausse2009} ($\chi_{\rm fit} = 0.369\pm 0.012$), and 
Ref.~\cite{Rezzolla:2007xa} ($\chi_{\rm fit} = 0.366 \pm 0.002$).
This result is a first step toward a better 
understanding of the relation between the properties of 
the remnant hole and those of the merging holes when the latter
are nearly extremal.
Numerical simulations that directly 
measure this relation (instead of extrapolating from lower-spin results) 
will yield greater understanding of the properties of 
black holes produced from merging extremal holes.

Figure~\ref{fig:orbits} shows the individual horizon trajectories and 
the real part of the $(\ell,m)=(2,2)$ 
spherical harmonic mode 
of the emitted gravitational waveform. We 
extract waves on a series of concentric spherical shells; the waveform 
shown was extracted on the outermost spherical shell 
(at radius $r=405 M$). 
Accurate gravitational waveforms obtained from this and 
future simulations with spins beyond the Bowen-York limit 
will be useful for calibrating analytic template banks for 
gravitational-wave searches.

The Christodoulou mass of the final black hole is 
$M_{\rm final}/M_{\rm relax}=0.9683\pm 0.0001$, 
where $M_{\rm relax}=1.0003 M$ is the sum of the masses of the 
individual holes after the initial relaxation 
and where the uncertainties 
of $M_{\rm final}$ and $M_{\rm relax}$ are estimated as 
the difference between the highest and second-highest resolution. 
Under the assumption that each hole has a constant mass $M_{\rm relax}$ 
throughout the inspiral (which holds within $O\left(10^{-5}\right)$ 
in our simulation after the holes have relaxed),
the quantity $1-M_{\rm final}/M_{\rm relax}$
represents the fraction
of the initial mass that would have been radiated from 
$t=-\infty$ to $t=+\infty$, 
had our simulation contained the entire inspiral 
instead of just the final 12.5 orbits.

\label{sec:Conclusion}

Our results demonstrate for the first time 
that it is possible to simulate 
merging black holes with spins larger than the 
Bowen-York limit of $\chi=0.93$, the highest 
spin previously obtainable.
Because 
astrophysical black holes may be nearly extremal, these simulations 
have astrophysical as well as physical relevance.
In particular, this work 
opens the way to use numerical simulations 
to explore the strong-field dynamics of merging, nearly extremal 
 black holes, to 
gain a better understanding of the properties of the remnant hole formed 
by a nearly extremal BBH merger, and 
to provide high-spin gravitational waveforms for data analysis.

\begin{acknowledgments}
We 
are pleased to thank 
Nick Taylor for a gauge modification that allows 
us to use the nonsmooth maps of Ref.~\cite{Szilagyi:2009qz} throughout 
our evolutions and
Larry Kidder, Robert Owen, Harald Pfeiffer, 
Saul Teukolsky, and Kip Thorne
for helpful discussions.
This work was supported in part by 
grants from the Sherman Fairchild Foundation to Caltech and Cornell 
and from the Brinson Foundation to Caltech; by NSF Grants No. PHY-0601459 and 
No. PHY-1005655 at Caltech; by NASA Grant No. NNX09AF97G at Caltech; 
by NSF Grants No. PHY-0969111 and No. PHY-1005426 at Cornell; 
and by NASA Grant No. NNX09AF96G at Cornell. 
The numerical computations presented in this paper were performed primarily 
on the Caltech compute cluster {\sc zwicky}, which was cofunded by the 
Sherman Fairchild Foundation. Some 
computations were also performed on the GPC supercomputer at the SciNet HPC 
Consortium; SciNet is funded by: the Canada Foundation for Innovation under 
the auspices of Compute Canada; the Government of Ontario; 
Ontario Research Fund - Research Excellence; and the University of Toronto. 
Some computations were performed in part using TeraGrid resources 
provided by NCSA's Ranger cluster under Grant No. TG-PHY990007N.
\end{acknowledgments}



\bibliography{References/References}

\begin{thebibliography}{50}%
\makeatletter
\providecommand \@ifxundefined [1]{%
 \@ifx{#1\undefined}
}%
\providecommand \@ifnum [1]{%
 \ifnum #1\expandafter \@firstoftwo
 \else \expandafter \@secondoftwo
 \fi
}%
\providecommand \@ifx [1]{%
 \ifx #1\expandafter \@firstoftwo
 \else \expandafter \@secondoftwo
 \fi
}%
\providecommand \natexlab [1]{#1}%
\providecommand \enquote  [1]{``#1''}%
\providecommand \bibnamefont  [1]{#1}%
\providecommand \bibfnamefont [1]{#1}%
\providecommand \citenamefont [1]{#1}%
\providecommand \href@noop [0]{\@secondoftwo}%
\providecommand \href [0]{\begingroup \@sanitize@url \@href}%
\providecommand \@href[1]{\@@startlink{#1}\@@href}%
\providecommand \@@href[1]{\endgroup#1\@@endlink}%
\providecommand \@sanitize@url [0]{\catcode `\\12\catcode `\$12\catcode
  `\&12\catcode `\#12\catcode `\^12\catcode `\_12\catcode `\%12\relax}%
\providecommand \@@startlink[1]{}%
\providecommand \@@endlink[0]{}%
\providecommand \url  [0]{\begingroup\@sanitize@url \@url }%
\providecommand \@url [1]{\endgroup\@href {#1}{\urlprefix }}%
\providecommand \urlprefix  [0]{URL }%
\providecommand \Eprint [0]{\href }%
\providecommand \doibase [0]{http://dx.doi.org/}%
\providecommand \selectlanguage [0]{\@gobble}%
\providecommand \bibinfo  [0]{\@secondoftwo}%
\providecommand \bibfield  [0]{\@secondoftwo}%
\providecommand \translation [1]{[#1]}%
\providecommand \BibitemOpen [0]{}%
\providecommand \bibitemStop [0]{}%
\providecommand \bibitemNoStop [0]{.\EOS\space}%
\providecommand \EOS [0]{\spacefactor3000\relax}%
\providecommand \BibitemShut  [1]{\csname bibitem#1\endcsname}%
\let\auto@bib@innerbib\@empty
\bibitem [{\citenamefont {Tichy}\ and\ \citenamefont
  {Marronetti}(2008)}]{Tichy2008}%
  \BibitemOpen
  \bibfield  {author} {\bibinfo {author} {\bibfnamefont {W.}~\bibnamefont
  {Tichy}}\ and\ \bibinfo {author} {\bibfnamefont {P.}~\bibnamefont
  {Marronetti}},\ }\href@noop {} {\bibfield  {journal} {\bibinfo  {journal}
  {Phys. Rev. D}\ }\textbf {\bibinfo {volume} {78}},\ \bibinfo {pages}
  {081501(R)} (\bibinfo {year} {2008})},\ \Eprint
  {http://arxiv.org/abs/arXiv:0807.2985 [gr-qc]} {arXiv:0807.2985 [gr-qc]}
  \BibitemShut {NoStop}%
\bibitem [{\citenamefont {{U. Sperhake, E. Berti, V. Cardoso, J. Gonz\'alez, B.
  Br\"ugmann and M. Ansorg}}(2008)}]{Sperhake2008Ecc}%
  \BibitemOpen
  \bibfield  {author} {\bibinfo {author} {\bibnamefont {{U. Sperhake, E. Berti,
  V. Cardoso, J. Gonz\'alez, B. Br\"ugmann and M. Ansorg}}},\ }\href@noop {}
  {\bibfield  {journal} {\bibinfo  {journal} {Phys.\ Rev.\ D}\ }\textbf
  {\bibinfo {volume} {78}},\ \bibinfo {pages} {064069} (\bibinfo {year}
  {2008})},\ \Eprint {http://arxiv.org/abs/arXiv:gr-qc/0710.3823}
  {arXiv:gr-qc/0710.3823} \BibitemShut {NoStop}%
\bibitem [{\citenamefont {Lousto}\ \emph {et~al.}(2010)\citenamefont {Lousto},
  \citenamefont {Campanelli}, \citenamefont {Zlochower},\ and\ \citenamefont
  {Nakano}}]{Lousto2009}%
  \BibitemOpen
  \bibfield  {author} {\bibinfo {author} {\bibfnamefont {C.~O.}\ \bibnamefont
  {Lousto}}, \bibinfo {author} {\bibfnamefont {M.}~\bibnamefont {Campanelli}},
  \bibinfo {author} {\bibfnamefont {Y.}~\bibnamefont {Zlochower}}, \ and\
  \bibinfo {author} {\bibfnamefont {H.}~\bibnamefont {Nakano}},\ }\href@noop {}
  {\bibfield  {journal} {\bibinfo  {journal} {Class. Quant. Grav.}\ }\textbf
  {\bibinfo {volume} {27}},\ \bibinfo {pages} {114006} (\bibinfo {year}
  {2010})},\ \Eprint {http://arxiv.org/abs/arXiv:0904.3541 [gr-qc]}
  {arXiv:0904.3541 [gr-qc]} \BibitemShut {NoStop}%
\bibitem [{\citenamefont {Volonteri}\ \emph {et~al.}(2005)\citenamefont
  {Volonteri}, \citenamefont {Madau}, \citenamefont {Quataert},\ and\
  \citenamefont {Rees}}]{VolonteriEtAl:2005}%
  \BibitemOpen
  \bibfield  {author} {\bibinfo {author} {\bibfnamefont {M.}~\bibnamefont
  {Volonteri}}, \bibinfo {author} {\bibfnamefont {P.}~\bibnamefont {Madau}},
  \bibinfo {author} {\bibfnamefont {E.}~\bibnamefont {Quataert}}, \ and\
  \bibinfo {author} {\bibfnamefont {M.~J.}\ \bibnamefont {Rees}},\ }\href@noop
  {} {\bibfield  {journal} {\bibinfo  {journal} {Astrophys.\ J.}\ }\textbf
  {\bibinfo {volume} {620}},\ \bibinfo {pages} {69} (\bibinfo {year}
  {2005})}\BibitemShut {NoStop}%
\bibitem [{\citenamefont {Berti}\ and\ \citenamefont
  {Volonteri}(2008)}]{BertiVolonteri:2008}%
  \BibitemOpen
  \bibfield  {author} {\bibinfo {author} {\bibfnamefont {E.}~\bibnamefont
  {Berti}}\ and\ \bibinfo {author} {\bibfnamefont {M.}~\bibnamefont
  {Volonteri}},\ }\href@noop {} {\bibfield  {journal} {\bibinfo  {journal}
  {Astrophys.\ J.}\ }\textbf {\bibinfo {volume} {684}},\ \bibinfo {pages} {822}
  (\bibinfo {year} {2008})},\ \Eprint {http://arxiv.org/abs/0802.0025v2}
  {arXiv:0802.0025v2 [gr-qc]} \BibitemShut {NoStop}%
\bibitem [{\citenamefont {Thorne}(1974)}]{Thorne:1974}%
  \BibitemOpen
  \bibfield  {author} {\bibinfo {author} {\bibfnamefont {K.~S.}\ \bibnamefont
  {Thorne}},\ }\href@noop {} {\bibfield  {journal} {\bibinfo  {journal}
  {Astrophys.\ J.}\ }\textbf {\bibinfo {volume} {191}},\ \bibinfo {pages} {507}
  (\bibinfo {year} {1974})}\BibitemShut {NoStop}%
\bibitem [{\citenamefont {Gammie}\ \emph {et~al.}(2004)\citenamefont {Gammie},
  \citenamefont {Shapiro},\ and\ \citenamefont {McKinney}}]{GammieEtAl:2004}%
  \BibitemOpen
  \bibfield  {author} {\bibinfo {author} {\bibfnamefont {C.~F.}\ \bibnamefont
  {Gammie}}, \bibinfo {author} {\bibfnamefont {S.~L.}\ \bibnamefont {Shapiro}},
  \ and\ \bibinfo {author} {\bibfnamefont {J.~C.}\ \bibnamefont {McKinney}},\
  }\href@noop {} {\bibfield  {journal} {\bibinfo  {journal} {Astrophys.\ J.}\
  }\textbf {\bibinfo {volume} {602}},\ \bibinfo {pages} {312} (\bibinfo {year}
  {2004})}\BibitemShut {NoStop}%
\bibitem [{\citenamefont {Shapiro}(2005)}]{Shapiro:2005}%
  \BibitemOpen
  \bibfield  {author} {\bibinfo {author} {\bibfnamefont {S.~L.}\ \bibnamefont
  {Shapiro}},\ }\href@noop {} {\bibfield  {journal} {\bibinfo  {journal}
  {Astrophys.\ J.}\ }\textbf {\bibinfo {volume} {620}},\ \bibinfo {pages} {59}
  (\bibinfo {year} {2005})}\BibitemShut {NoStop}%
\bibitem [{\citenamefont {Rezzolla}\ \emph {et~al.}(2008)\citenamefont
  {Rezzolla} \emph {et~al.}}]{Rezzolla:2007xa}%
  \BibitemOpen
  \bibfield  {author} {\bibinfo {author} {\bibfnamefont {L.}~\bibnamefont
  {Rezzolla}} \emph {et~al.},\ }\href@noop {} {\bibfield  {journal} {\bibinfo
  {journal} {Astrophys.\ J.}\ }\textbf {\bibinfo {volume} {679}},\ \bibinfo
  {pages} {1422} (\bibinfo {year} {2008})},\ \Eprint
  {http://arxiv.org/abs/0708.3999} {arXiv:0708.3999 [gr-qc]} \BibitemShut
  {NoStop}%
\bibitem [{\citenamefont {Kesden}(2008)}]{Kesden2008}%
  \BibitemOpen
  \bibfield  {author} {\bibinfo {author} {\bibfnamefont {M.}~\bibnamefont
  {Kesden}},\ }\href@noop {} {\bibfield  {journal} {\bibinfo  {journal} {Phys.\
  Rev.\ D}\ }\textbf {\bibinfo {volume} {78}},\ \bibinfo {pages} {084030}
  (\bibinfo {year} {2008})}\BibitemShut {NoStop}%
\bibitem [{\citenamefont {Kesden}\ \emph {et~al.}(2010)\citenamefont {Kesden},
  \citenamefont {Lockhart},\ and\ \citenamefont {Phinney}}]{KesdenEtAl:2010}%
  \BibitemOpen
  \bibfield  {author} {\bibinfo {author} {\bibfnamefont {M.}~\bibnamefont
  {Kesden}}, \bibinfo {author} {\bibfnamefont {G.}~\bibnamefont {Lockhart}}, \
  and\ \bibinfo {author} {\bibfnamefont {E.~S.}\ \bibnamefont {Phinney}},\
  }\href@noop {} {\bibfield  {journal} {\bibinfo  {journal} {Phys.\ Rev.\ D}\
  }\textbf {\bibinfo {volume} {82}},\ \bibinfo {pages} {124045} (\bibinfo
  {year} {2010})},\ \Eprint {http://arxiv.org/abs/arXiv:1005.0627}
  {arXiv:1005.0627} \BibitemShut {NoStop}%
\bibitem [{\citenamefont {Wang}\ \emph {et~al.}(2006)\citenamefont {Wang},
  \citenamefont {Chen}, \citenamefont {Ho},\ and\ \citenamefont
  {McLure}}]{WangEtAl:2006}%
  \BibitemOpen
  \bibfield  {author} {\bibinfo {author} {\bibfnamefont {J.-M.}\ \bibnamefont
  {Wang}}, \bibinfo {author} {\bibfnamefont {Y.-M.}\ \bibnamefont {Chen}},
  \bibinfo {author} {\bibfnamefont {L.~C.}\ \bibnamefont {Ho}}, \ and\ \bibinfo
  {author} {\bibfnamefont {R.~J.}\ \bibnamefont {McLure}},\ }\href@noop {}
  {\bibfield  {journal} {\bibinfo  {journal} {Astrophys.\ J.}\ }\textbf
  {\bibinfo {volume} {642}},\ \bibinfo {pages} {L111} (\bibinfo {year}
  {2006})}\BibitemShut {NoStop}%
\bibitem [{\citenamefont {McClintock}\ \emph {et~al.}(2006)\citenamefont
  {McClintock}, \citenamefont {Shafee}, \citenamefont {Narayan}, \citenamefont
  {Remillard}, \citenamefont {Davis},\ and\ \citenamefont
  {Li}}]{McClintockEtAl:2006}%
  \BibitemOpen
  \bibfield  {author} {\bibinfo {author} {\bibfnamefont {J.~E.}\ \bibnamefont
  {McClintock}}, \bibinfo {author} {\bibfnamefont {R.}~\bibnamefont {Shafee}},
  \bibinfo {author} {\bibfnamefont {R.}~\bibnamefont {Narayan}}, \bibinfo
  {author} {\bibfnamefont {R.~A.}\ \bibnamefont {Remillard}}, \bibinfo {author}
  {\bibfnamefont {S.~W.}\ \bibnamefont {Davis}}, \ and\ \bibinfo {author}
  {\bibfnamefont {L.-X.}\ \bibnamefont {Li}},\ }\href@noop {} {\bibfield
  {journal} {\bibinfo  {journal} {Astrophys.\ J.}\ }\textbf {\bibinfo {volume}
  {652}},\ \bibinfo {pages} {518} (\bibinfo {year} {2006})}\BibitemShut
  {NoStop}%
\bibitem [{\citenamefont {Middleton}\ \emph {et~al.}(2006)\citenamefont
  {Middleton}, \citenamefont {Done}, \citenamefont {Gierli\'{n}ski},\ and\
  \citenamefont {Davis}}]{Middleton:2006}%
  \BibitemOpen
  \bibfield  {author} {\bibinfo {author} {\bibfnamefont {M.}~\bibnamefont
  {Middleton}}, \bibinfo {author} {\bibfnamefont {C.}~\bibnamefont {Done}},
  \bibinfo {author} {\bibfnamefont {M.}~\bibnamefont {Gierli\'{n}ski}}, \ and\
  \bibinfo {author} {\bibfnamefont {S.~W.}\ \bibnamefont {Davis}},\ }\href@noop
  {} {\bibfield  {journal} {\bibinfo  {journal} {Mon. Not. R. Astron. Soc.}\
  }\textbf {\bibinfo {volume} {373}},\ \bibinfo {pages} {1004} (\bibinfo {year}
  {2006})}\BibitemShut {NoStop}%
\bibitem [{\citenamefont {Blum}\ \emph {et~al.}(2009)\citenamefont {Blum},
  \citenamefont {Miller}, \citenamefont {Fabian}, \citenamefont {Miller},
  \citenamefont {Homan}, \citenamefont {van~der Klis}, \citenamefont
  {Cackett},\ and\ \citenamefont {Reis}}]{Blum:2009}%
  \BibitemOpen
  \bibfield  {author} {\bibinfo {author} {\bibfnamefont {J.~L.}\ \bibnamefont
  {Blum}}, \bibinfo {author} {\bibfnamefont {J.~M.}\ \bibnamefont {Miller}},
  \bibinfo {author} {\bibfnamefont {A.~C.}\ \bibnamefont {Fabian}}, \bibinfo
  {author} {\bibfnamefont {M.~C.}\ \bibnamefont {Miller}}, \bibinfo {author}
  {\bibfnamefont {J.}~\bibnamefont {Homan}}, \bibinfo {author} {\bibfnamefont
  {M.}~\bibnamefont {van~der Klis}}, \bibinfo {author} {\bibfnamefont {E.~M.}\
  \bibnamefont {Cackett}}, \ and\ \bibinfo {author} {\bibfnamefont {R.~C.}\
  \bibnamefont {Reis}},\ }\href@noop {} {\bibfield  {journal} {\bibinfo
  {journal} {Astrophys.\ J.}\ }\textbf {\bibinfo {volume} {706}},\ \bibinfo
  {pages} {60} (\bibinfo {year} {2009})}\BibitemShut {NoStop}%
\bibitem [{\citenamefont {Pretorius}(2005)}]{Pretorius2005a}%
  \BibitemOpen
  \bibfield  {author} {\bibinfo {author} {\bibfnamefont {F.}~\bibnamefont
  {Pretorius}},\ }\href@noop {} {\bibfield  {journal} {\bibinfo  {journal}
  {Phys.\ Rev.\ Lett.}\ }\textbf {\bibinfo {volume} {95}},\ \bibinfo {eid}
  {121101} (\bibinfo {year} {2005})}\BibitemShut {NoStop}%
\bibitem [{\citenamefont {Brandt}\ and\ \citenamefont
  {Br{\"u}gmann}(1997)}]{Brandt1997}%
  \BibitemOpen
  \bibfield  {author} {\bibinfo {author} {\bibfnamefont {S.}~\bibnamefont
  {Brandt}}\ and\ \bibinfo {author} {\bibfnamefont {B.}~\bibnamefont
  {Br{\"u}gmann}},\ }\href@noop {} {\bibfield  {journal} {\bibinfo  {journal}
  {Phys.\ Rev.\ Lett.}\ }\textbf {\bibinfo {volume} {78}},\ \bibinfo {pages}
  {3606} (\bibinfo {year} {1997})}\BibitemShut {NoStop}%
\bibitem [{\citenamefont {Bowen}(1979)}]{bowen79}%
  \BibitemOpen
  \bibfield  {author} {\bibinfo {author} {\bibfnamefont {J.~M.}\ \bibnamefont
  {Bowen}},\ }\href@noop {} {\bibfield  {journal} {\bibinfo  {journal} {Gen.\
  Relativ.\ Gravit.}\ }\textbf {\bibinfo {volume} {11}},\ \bibinfo {pages}
  {227} (\bibinfo {year} {1979})}\BibitemShut {NoStop}%
\bibitem [{\citenamefont {Bowen}\ and\ \citenamefont {{York,
  Jr.}}(1980)}]{Bowen-York:1980}%
  \BibitemOpen
  \bibfield  {author} {\bibinfo {author} {\bibfnamefont {J.~M.}\ \bibnamefont
  {Bowen}}\ and\ \bibinfo {author} {\bibfnamefont {J.~W.}\ \bibnamefont {{York,
  Jr.}}},\ }\href@noop {} {\bibfield  {journal} {\bibinfo  {journal} {Phys.\
  Rev.\ D}\ }\textbf {\bibinfo {volume} {21}},\ \bibinfo {pages} {2047}
  (\bibinfo {year} {1980})}\BibitemShut {NoStop}%
\bibitem [{\citenamefont {{York, Jr.}}(1980)}]{york80}%
  \BibitemOpen
  \bibfield  {author} {\bibinfo {author} {\bibfnamefont {J.~W.}\ \bibnamefont
  {{York, Jr.}}},\ }in\ \href@noop {} {\emph {\bibinfo {booktitle} {Essays in
  General Relativity}}},\ \bibinfo {editor} {edited by\ \bibinfo {editor}
  {\bibfnamefont {F.~J.}\ \bibnamefont {Tipler}}}\ (\bibinfo  {publisher}
  {Academic},\ \bibinfo {address} {New York},\ \bibinfo {year} {1980})\ pp.\
  \bibinfo {pages} {39--58}\BibitemShut {NoStop}%
\bibitem [{\citenamefont {Garat}\ and\ \citenamefont
  {Price}(2000)}]{GaratPrice:2000}%
  \BibitemOpen
  \bibfield  {author} {\bibinfo {author} {\bibfnamefont {A.}~\bibnamefont
  {Garat}}\ and\ \bibinfo {author} {\bibfnamefont {R.~H.}\ \bibnamefont
  {Price}},\ }\href@noop {} {\bibfield  {journal} {\bibinfo  {journal} {Phys.\
  Rev.\ D}\ }\textbf {\bibinfo {volume} {61}},\ \bibinfo {pages} {124011}
  (\bibinfo {year} {2000})}\BibitemShut {NoStop}%
\bibitem [{\citenamefont {{Valiente Kroon}}(2004)}]{Kroon:2004}%
  \BibitemOpen
  \bibfield  {author} {\bibinfo {author} {\bibfnamefont {J.~A.}\ \bibnamefont
  {{Valiente Kroon}}},\ }\href@noop {} {\bibfield  {journal} {\bibinfo
  {journal} {Phys. Rev. Lett.}\ }\textbf {\bibinfo {volume} {92}},\ \bibinfo
  {pages} {041101} (\bibinfo {year} {2004})}\BibitemShut {NoStop}%
\bibitem [{\citenamefont {Cook}\ and\ \citenamefont {{York,
  Jr.}}(1990)}]{cook90}%
  \BibitemOpen
  \bibfield  {author} {\bibinfo {author} {\bibfnamefont {G.~B.}\ \bibnamefont
  {Cook}}\ and\ \bibinfo {author} {\bibfnamefont {J.~W.}\ \bibnamefont {{York,
  Jr.}}},\ }\href@noop {} {\bibfield  {journal} {\bibinfo  {journal} {Phys.\
  Rev.\ D}\ }\textbf {\bibinfo {volume} {41}},\ \bibinfo {pages} {1077}
  (\bibinfo {year} {1990})}\BibitemShut {NoStop}%
\bibitem [{\citenamefont {Dain}\ \emph {et~al.}(2002)\citenamefont {Dain},
  \citenamefont {Lousto},\ and\ \citenamefont {Takahashi}}]{DainEtAl:2002}%
  \BibitemOpen
  \bibfield  {author} {\bibinfo {author} {\bibfnamefont {S.}~\bibnamefont
  {Dain}}, \bibinfo {author} {\bibfnamefont {C.~O.}\ \bibnamefont {Lousto}}, \
  and\ \bibinfo {author} {\bibfnamefont {R.}~\bibnamefont {Takahashi}},\
  }\href@noop {} {\bibfield  {journal} {\bibinfo  {journal} {Phys. Rev. D}\
  }\textbf {\bibinfo {volume} {65}},\ \bibinfo {pages} {104038} (\bibinfo
  {year} {2002})}\BibitemShut {NoStop}%
\bibitem [{\citenamefont {Hannam}\ \emph {et~al.}(2009)\citenamefont {Hannam},
  \citenamefont {Husa},\ and\ \citenamefont {Murchadha}}]{HannamEtAl:2009}%
  \BibitemOpen
  \bibfield  {author} {\bibinfo {author} {\bibfnamefont {M.}~\bibnamefont
  {Hannam}}, \bibinfo {author} {\bibfnamefont {S.}~\bibnamefont {Husa}}, \ and\
  \bibinfo {author} {\bibfnamefont {N.~O.}\ \bibnamefont {Murchadha}},\
  }\href@noop {} {\bibfield  {journal} {\bibinfo  {journal} {Phys.\ Rev.\ D}\
  }\textbf {\bibinfo {volume} {80}},\ \bibinfo {pages} {124007} (\bibinfo
  {year} {2009})}\BibitemShut {NoStop}%
\bibitem [{\citenamefont {Hannam}\ \emph {et~al.}(2010)\citenamefont {Hannam},
  \citenamefont {Husa}, \citenamefont {Ohme}, \citenamefont {M\"{u}ller},\ and\
  \citenamefont {Br\"{u}gmann}}]{HannamEtAl:2010}%
  \BibitemOpen
  \bibfield  {author} {\bibinfo {author} {\bibfnamefont {M.}~\bibnamefont
  {Hannam}}, \bibinfo {author} {\bibfnamefont {S.}~\bibnamefont {Husa}},
  \bibinfo {author} {\bibfnamefont {F.}~\bibnamefont {Ohme}}, \bibinfo {author}
  {\bibfnamefont {D.}~\bibnamefont {M\"{u}ller}}, \ and\ \bibinfo {author}
  {\bibfnamefont {B.}~\bibnamefont {Br\"{u}gmann}},\ }\href@noop {} {\bibfield
  {journal} {\bibinfo  {journal} {Phys.\ Rev.\ D}\ }\textbf {\bibinfo {volume}
  {82}},\ \bibinfo {pages} {124008} (\bibinfo {year} {2010})},\ \Eprint
  {http://arxiv.org/abs/arXiv:1007.4789} {arXiv:1007.4789} \BibitemShut
  {NoStop}%
\bibitem [{\citenamefont {Marronetti}\ \emph {et~al.}(2008)\citenamefont
  {Marronetti}, \citenamefont {Tichy}, \citenamefont {Br{\"{u}}gmann},
  \citenamefont {Gonz{\'{a}}lez},\ and\ \citenamefont
  {Sperhake}}]{MarronettiEtal:2008}%
  \BibitemOpen
  \bibfield  {author} {\bibinfo {author} {\bibfnamefont {P.}~\bibnamefont
  {Marronetti}}, \bibinfo {author} {\bibfnamefont {W.}~\bibnamefont {Tichy}},
  \bibinfo {author} {\bibfnamefont {B.}~\bibnamefont {Br{\"{u}}gmann}},
  \bibinfo {author} {\bibfnamefont {J.}~\bibnamefont {Gonz{\'{a}}lez}}, \ and\
  \bibinfo {author} {\bibfnamefont {U.}~\bibnamefont {Sperhake}},\ }\href@noop
  {} {\bibfield  {journal} {\bibinfo  {journal} {Phys.\ Rev.\ D}\ }\textbf
  {\bibinfo {volume} {77}},\ \bibinfo {pages} {064010} (\bibinfo {year}
  {2008})}\BibitemShut {NoStop}%
\bibitem [{\citenamefont {Dain}\ \emph {et~al.}(2008)\citenamefont {Dain},
  \citenamefont {Lousto},\ and\ \citenamefont {Zlochower}}]{DainEtAl:2008}%
  \BibitemOpen
  \bibfield  {author} {\bibinfo {author} {\bibfnamefont {S.}~\bibnamefont
  {Dain}}, \bibinfo {author} {\bibfnamefont {C.~O.}\ \bibnamefont {Lousto}}, \
  and\ \bibinfo {author} {\bibfnamefont {Y.}~\bibnamefont {Zlochower}},\
  }\href@noop {} {\bibfield  {journal} {\bibinfo  {journal} {Phys.\ Rev.\ D}\
  }\textbf {\bibinfo {volume} {78}},\ \bibinfo {pages} {024039} (\bibinfo
  {year} {2008})},\ \Eprint {http://arxiv.org/abs/0803.0351v2}
  {arXiv:0803.0351v2 [gr-qc]} \BibitemShut {NoStop}%
\bibitem [{\citenamefont {Liu}\ \emph {et~al.}(2009)\citenamefont {Liu},
  \citenamefont {Etienne},\ and\ \citenamefont {Shapiro}}]{Liu:2009}%
  \BibitemOpen
  \bibfield  {author} {\bibinfo {author} {\bibfnamefont {Y.~T.}\ \bibnamefont
  {Liu}}, \bibinfo {author} {\bibfnamefont {Z.~B.}\ \bibnamefont {Etienne}}, \
  and\ \bibinfo {author} {\bibfnamefont {S.~L.}\ \bibnamefont {Shapiro}},\
  }\href@noop {} {\bibfield  {journal} {\bibinfo  {journal} {Phys.\ Rev.\ D}\
  }\textbf {\bibinfo {volume} {80}},\ \bibinfo {pages} {121503(R)} (\bibinfo
  {year} {2009})}\BibitemShut {NoStop}%
\bibitem [{\citenamefont {Brandt}\ and\ \citenamefont
  {Seidel}(1995)}]{BrandtSeidel:1995a}%
  \BibitemOpen
  \bibfield  {author} {\bibinfo {author} {\bibfnamefont {S.~R.}\ \bibnamefont
  {Brandt}}\ and\ \bibinfo {author} {\bibfnamefont {E.}~\bibnamefont
  {Seidel}},\ }\href@noop {} {\bibfield  {journal} {\bibinfo  {journal} {Phys.\
  Rev.\ D}\ }\textbf {\bibinfo {volume} {52}},\ \bibinfo {pages} {856}
  (\bibinfo {year} {1995})}\BibitemShut {NoStop}%
\bibitem [{\citenamefont {Brandt}\ and\ \citenamefont
  {Seidel}(1996)}]{BrandtSeidel:1996}%
  \BibitemOpen
  \bibfield  {author} {\bibinfo {author} {\bibfnamefont {S.~R.}\ \bibnamefont
  {Brandt}}\ and\ \bibinfo {author} {\bibfnamefont {E.}~\bibnamefont
  {Seidel}},\ }\href@noop {} {\ \textbf {\bibinfo {volume} {54}},\ \bibinfo
  {pages} {1403} (\bibinfo {year} {1996})}\BibitemShut {NoStop}%
\bibitem [{\citenamefont {Hannam}\ \emph {et~al.}(2007)\citenamefont {Hannam},
  \citenamefont {Husa}, \citenamefont {Br{\"u}gmann}, \citenamefont
  {Gonzalez},\ and\ \citenamefont {Sperhake}}]{Hannam2007b}%
  \BibitemOpen
  \bibfield  {author} {\bibinfo {author} {\bibfnamefont {M.}~\bibnamefont
  {Hannam}}, \bibinfo {author} {\bibfnamefont {S.}~\bibnamefont {Husa}},
  \bibinfo {author} {\bibfnamefont {B.}~\bibnamefont {Br{\"u}gmann}}, \bibinfo
  {author} {\bibfnamefont {J.~A.}\ \bibnamefont {Gonzalez}}, \ and\ \bibinfo
  {author} {\bibfnamefont {U.}~\bibnamefont {Sperhake}},\ }\href@noop {}
  {\bibfield  {journal} {\bibinfo  {journal} {Class.\ Quantum Grav.}\ }\textbf
  {\bibinfo {volume} {24}},\ \bibinfo {pages} {S15} (\bibinfo {year} {2007})},\
  \Eprint {http://arxiv.org/abs/gr-qc/0612001} {gr-qc/0612001} \BibitemShut
  {NoStop}%
\bibitem [{\citenamefont {Dain}(2001{\natexlab{a}})}]{Dain:2001a}%
  \BibitemOpen
  \bibfield  {author} {\bibinfo {author} {\bibfnamefont {S.}~\bibnamefont
  {Dain}},\ }\href@noop {} {\bibfield  {journal} {\bibinfo  {journal} {Phys.\
  Rev.\ Lett.}\ }\textbf {\bibinfo {volume} {87}},\ \bibinfo {pages} {121102}
  (\bibinfo {year} {2001}{\natexlab{a}})}\BibitemShut {NoStop}%
\bibitem [{\citenamefont {Dain}(2001{\natexlab{b}})}]{Dain:2001b}%
  \BibitemOpen
  \bibfield  {author} {\bibinfo {author} {\bibfnamefont {S.}~\bibnamefont
  {Dain}},\ }\href@noop {} {\bibfield  {journal} {\bibinfo  {journal} {Phys.\
  Rev.\ D}\ }\textbf {\bibinfo {volume} {64}},\ \bibinfo {pages} {124002}
  (\bibinfo {year} {2001}{\natexlab{b}})}\BibitemShut {NoStop}%
\bibitem [{\citenamefont {Lovelace}\ \emph {et~al.}(2008)\citenamefont
  {Lovelace}, \citenamefont {Owen}, \citenamefont {Pfeiffer},\ and\
  \citenamefont {Chu}}]{Lovelace2008}%
  \BibitemOpen
  \bibfield  {author} {\bibinfo {author} {\bibfnamefont {G.}~\bibnamefont
  {Lovelace}}, \bibinfo {author} {\bibfnamefont {R.}~\bibnamefont {Owen}},
  \bibinfo {author} {\bibfnamefont {H.~P.}\ \bibnamefont {Pfeiffer}}, \ and\
  \bibinfo {author} {\bibfnamefont {T.}~\bibnamefont {Chu}},\ }\href@noop {}
  {\bibfield  {journal} {\bibinfo  {journal} {Phys.\ Rev.\ D}\ }\textbf
  {\bibinfo {volume} {78}},\ \bibinfo {pages} {084017} (\bibinfo {year}
  {2008})}\BibitemShut {NoStop}%
\bibitem [{\citenamefont {York}(1999)}]{York1999}%
  \BibitemOpen
  \bibfield  {author} {\bibinfo {author} {\bibfnamefont {J.~W.}\ \bibnamefont
  {York}},\ }\href {\doibase 10.1103/PhysRevLett.82.1350} {\bibfield  {journal}
  {\bibinfo  {journal} {Phys.\ Rev.\ Lett.}\ }\textbf {\bibinfo {volume}
  {82}},\ \bibinfo {pages} {1350} (\bibinfo {year} {1999})}\BibitemShut
  {NoStop}%
\bibitem [{\citenamefont {Cook}(2002)}]{Cook2002}%
  \BibitemOpen
  \bibfield  {author} {\bibinfo {author} {\bibfnamefont {G.~B.}\ \bibnamefont
  {Cook}},\ }\href {\doibase 10.1103/PhysRevD.65.084003} {\bibfield  {journal}
  {\bibinfo  {journal} {Phys.\ Rev.\ D}\ }\textbf {\bibinfo {volume} {65}},\
  \bibinfo {pages} {084003} (\bibinfo {year} {2002})}\BibitemShut {NoStop}%
\bibitem [{\citenamefont {Cook}\ and\ \citenamefont
  {Pfeiffer}(2004)}]{Cook2004}%
  \BibitemOpen
  \bibfield  {author} {\bibinfo {author} {\bibfnamefont {G.~B.}\ \bibnamefont
  {Cook}}\ and\ \bibinfo {author} {\bibfnamefont {H.~P.}\ \bibnamefont
  {Pfeiffer}},\ }\href {\doibase 10.1103/PhysRevD.70.104016} {\bibfield
  {journal} {\bibinfo  {journal} {Phys.\ Rev.\ D}\ }\textbf {\bibinfo {volume}
  {70}},\ \bibinfo {pages} {104016} (\bibinfo {year} {2004})}\BibitemShut
  {NoStop}%
\bibitem [{\citenamefont {{Caudill}}\ \emph {et~al.}(2006)\citenamefont
  {{Caudill}}, \citenamefont {{Cook}}, \citenamefont {{Grigsby}},\ and\
  \citenamefont {{Pfeiffer}}}]{Caudill-etal:2006}%
  \BibitemOpen
  \bibfield  {author} {\bibinfo {author} {\bibfnamefont {M.}~\bibnamefont
  {{Caudill}}}, \bibinfo {author} {\bibfnamefont {G.~B.}\ \bibnamefont
  {{Cook}}}, \bibinfo {author} {\bibfnamefont {J.~D.}\ \bibnamefont
  {{Grigsby}}}, \ and\ \bibinfo {author} {\bibfnamefont {H.~P.}\ \bibnamefont
  {{Pfeiffer}}},\ }\href@noop {} {\bibfield  {journal} {\bibinfo  {journal}
  {Phys.\ Rev.\ D}\ }\textbf {\bibinfo {volume} {74}},\ \bibinfo {pages}
  {064011} (\bibinfo {year} {2006})}\BibitemShut {NoStop}%
\bibitem [{\citenamefont {Gourgoulhon}\ \emph {et~al.}(2002)\citenamefont
  {Gourgoulhon}, \citenamefont {Grandcl{\'e}ment},\ and\ \citenamefont
  {Bonazzola}}]{Gourgoulhon2001}%
  \BibitemOpen
  \bibfield  {author} {\bibinfo {author} {\bibfnamefont {E.}~\bibnamefont
  {Gourgoulhon}}, \bibinfo {author} {\bibfnamefont {P.}~\bibnamefont
  {Grandcl{\'e}ment}}, \ and\ \bibinfo {author} {\bibfnamefont
  {S.}~\bibnamefont {Bonazzola}},\ }\href@noop {} {\bibfield  {journal}
  {\bibinfo  {journal} {Phys.\ Rev.\ D}\ }\textbf {\bibinfo {volume} {65}},\
  \bibinfo {pages} {044020} (\bibinfo {year} {2002})}\BibitemShut {NoStop}%
\bibitem [{\citenamefont {Grandcl\'ement}\ \emph {et~al.}(2002)\citenamefont
  {Grandcl\'ement}, \citenamefont {Gourgoulhon},\ and\ \citenamefont
  {Bonazzola}}]{Grandclement2002}%
  \BibitemOpen
  \bibfield  {author} {\bibinfo {author} {\bibfnamefont {P.}~\bibnamefont
  {Grandcl\'ement}}, \bibinfo {author} {\bibfnamefont {E.}~\bibnamefont
  {Gourgoulhon}}, \ and\ \bibinfo {author} {\bibfnamefont {S.}~\bibnamefont
  {Bonazzola}},\ }\href@noop {} {\bibfield  {journal} {\bibinfo  {journal}
  {Phys.\ Rev.\ D}\ }\textbf {\bibinfo {volume} {65}},\ \bibinfo {pages}
  {044021} (\bibinfo {year} {2002})}\BibitemShut {NoStop}%
\bibitem [{\citenamefont {Pfeiffer}\ \emph {et~al.}(2003)\citenamefont
  {Pfeiffer}, \citenamefont {Kidder}, \citenamefont {Scheel},\ and\
  \citenamefont {Teukolsky}}]{Pfeiffer2003}%
  \BibitemOpen
  \bibfield  {author} {\bibinfo {author} {\bibfnamefont {H.~P.}\ \bibnamefont
  {Pfeiffer}}, \bibinfo {author} {\bibfnamefont {L.~E.}\ \bibnamefont
  {Kidder}}, \bibinfo {author} {\bibfnamefont {M.~A.}\ \bibnamefont {Scheel}},
  \ and\ \bibinfo {author} {\bibfnamefont {S.~A.}\ \bibnamefont {Teukolsky}},\
  }\href@noop {} {\bibfield  {journal} {\bibinfo  {journal} {Comput.\ Phys.\
  Commun.}\ }\textbf {\bibinfo {volume} {152}},\ \bibinfo {pages} {253}
  (\bibinfo {year} {2003})}\BibitemShut {NoStop}%
\bibitem [{\citenamefont {Cook}\ and\ \citenamefont
  {Whiting}(2007)}]{Cook2007}%
  \BibitemOpen
  \bibfield  {author} {\bibinfo {author} {\bibfnamefont {G.~B.}\ \bibnamefont
  {Cook}}\ and\ \bibinfo {author} {\bibfnamefont {B.~F.}\ \bibnamefont
  {Whiting}},\ }\href {\doibase 10.1103/PhysRevD.76.041501} {\bibfield
  {journal} {\bibinfo  {journal} {Phys.\ Rev.\ D}\ }\textbf {\bibinfo {volume}
  {76}},\ \bibinfo {eid} {041501(R)} (\bibinfo {year} {2007})}\BibitemShut
  {NoStop}%
\bibitem [{\citenamefont {Buonanno}\ \emph {et~al.}(2010)\citenamefont
  {Buonanno}, \citenamefont {Kidder}, \citenamefont {Mrou\'{e}}, \citenamefont
  {Pfeiffer},\ and\ \citenamefont {Taracchini}}]{BuonannoEtAl:2010}%
  \BibitemOpen
  \bibfield  {author} {\bibinfo {author} {\bibfnamefont {A.}~\bibnamefont
  {Buonanno}}, \bibinfo {author} {\bibfnamefont {L.~E.}\ \bibnamefont
  {Kidder}}, \bibinfo {author} {\bibfnamefont {A.~H.}\ \bibnamefont
  {Mrou\'{e}}}, \bibinfo {author} {\bibfnamefont {H.~P.}\ \bibnamefont
  {Pfeiffer}}, \ and\ \bibinfo {author} {\bibfnamefont {A.}~\bibnamefont
  {Taracchini}},\ }\href@noop {} {\  (\bibinfo {year} {2010})},\ \bibinfo
  {note} {submitted to Phys. Rev. D},\ \Eprint
  {http://arxiv.org/abs/arXiv:1012.1549} {arXiv:1012.1549} \BibitemShut
  {NoStop}%
\bibitem [{\citenamefont {Pfeiffer}\ \emph {et~al.}(2007)\citenamefont
  {Pfeiffer}, \citenamefont {Brown}, \citenamefont {Kidder}, \citenamefont
  {Lindblom}, \citenamefont {Lovelace},\ and\ \citenamefont
  {Scheel}}]{Pfeiffer-Brown-etal:2007}%
  \BibitemOpen
  \bibfield  {author} {\bibinfo {author} {\bibfnamefont {H.~P.}\ \bibnamefont
  {Pfeiffer}}, \bibinfo {author} {\bibfnamefont {D.~A.}\ \bibnamefont {Brown}},
  \bibinfo {author} {\bibfnamefont {L.~E.}\ \bibnamefont {Kidder}}, \bibinfo
  {author} {\bibfnamefont {L.}~\bibnamefont {Lindblom}}, \bibinfo {author}
  {\bibfnamefont {G.}~\bibnamefont {Lovelace}}, \ and\ \bibinfo {author}
  {\bibfnamefont {M.~A.}\ \bibnamefont {Scheel}},\ }\href@noop {} {\bibfield
  {journal} {\bibinfo  {journal} {Class.\ Quantum Grav.}\ }\textbf {\bibinfo
  {volume} {24}},\ \bibinfo {pages} {S59} (\bibinfo {year} {2007})}\BibitemShut
  {NoStop}%
\bibitem [{SpE()}]{SpECwebsite}%
  \BibitemOpen
  \href@noop {} {}\bibinfo {howpublished}
  {\url{http://www.black-holes.org/SpEC.html}}\BibitemShut {NoStop}%
\bibitem [{\citenamefont {Szilagyi}\ \emph {et~al.}(2009)\citenamefont
  {Szilagyi}, \citenamefont {Lindblom},\ and\ \citenamefont
  {Scheel}}]{Szilagyi:2009qz}%
  \BibitemOpen
  \bibfield  {author} {\bibinfo {author} {\bibfnamefont {B.}~\bibnamefont
  {Szilagyi}}, \bibinfo {author} {\bibfnamefont {L.}~\bibnamefont {Lindblom}},
  \ and\ \bibinfo {author} {\bibfnamefont {M.~A.}\ \bibnamefont {Scheel}},\
  }\href@noop {} {\bibfield  {journal} {\bibinfo  {journal} {Phys.\ Rev.\ D}\
  }\textbf {\bibinfo {volume} {80}},\ \bibinfo {pages} {124010} (\bibinfo
  {year} {2009})},\ \Eprint {http://arxiv.org/abs/0909.3557} {arXiv:0909.3557
  [gr-qc]} \BibitemShut {NoStop}%
\bibitem [{\citenamefont {Scheel}\ \emph {et~al.}(2006)\citenamefont {Scheel},
  \citenamefont {Pfeiffer}, \citenamefont {Lindblom}, \citenamefont {Kidder},
  \citenamefont {Rinne},\ and\ \citenamefont {Teukolsky}}]{Scheel2006}%
  \BibitemOpen
  \bibfield  {author} {\bibinfo {author} {\bibfnamefont {M.~A.}\ \bibnamefont
  {Scheel}}, \bibinfo {author} {\bibfnamefont {H.~P.}\ \bibnamefont
  {Pfeiffer}}, \bibinfo {author} {\bibfnamefont {L.}~\bibnamefont {Lindblom}},
  \bibinfo {author} {\bibfnamefont {L.~E.}\ \bibnamefont {Kidder}}, \bibinfo
  {author} {\bibfnamefont {O.}~\bibnamefont {Rinne}}, \ and\ \bibinfo {author}
  {\bibfnamefont {S.~A.}\ \bibnamefont {Teukolsky}},\ }\href@noop {} {\bibfield
   {journal} {\bibinfo  {journal} {Phys.\ Rev.\ D}\ }\textbf {\bibinfo {volume}
  {74}},\ \bibinfo {pages} {104006} (\bibinfo {year} {2006})}\BibitemShut
  {NoStop}%
\bibitem [{\citenamefont {Campanelli}\ \emph {et~al.}(2006)\citenamefont
  {Campanelli}, \citenamefont {Lousto},\ and\ \citenamefont
  {Zlochower}}]{Campanelli2006c}%
  \BibitemOpen
  \bibfield  {author} {\bibinfo {author} {\bibfnamefont {M.}~\bibnamefont
  {Campanelli}}, \bibinfo {author} {\bibfnamefont {C.~O.}\ \bibnamefont
  {Lousto}}, \ and\ \bibinfo {author} {\bibfnamefont {Y.}~\bibnamefont
  {Zlochower}},\ }\href@noop {} {\bibfield  {journal} {\bibinfo  {journal}
  {Phys.\ Rev.\ D}\ }\textbf {\bibinfo {volume} {74}},\ \bibinfo {pages}
  {041501(R)} (\bibinfo {year} {2006})},\ \Eprint
  {http://arxiv.org/abs/gr-qc/0604012} {gr-qc/0604012} \BibitemShut {NoStop}%
\bibitem [{\citenamefont {Barausse}\ and\ \citenamefont
  {Rezzolla}(2009)}]{Barausse2009}%
  \BibitemOpen
  \bibfield  {author} {\bibinfo {author} {\bibfnamefont {E.}~\bibnamefont
  {Barausse}}\ and\ \bibinfo {author} {\bibfnamefont {L.}~\bibnamefont
  {Rezzolla}},\ }\href@noop {} {\bibfield  {journal} {\bibinfo  {journal}
  {Astrophys.\ J.\ Lett.}\ }\textbf {\bibinfo {volume} {704}},\ \bibinfo
  {pages} {L40} (\bibinfo {year} {2009})},\ \Eprint
  {http://arxiv.org/abs/arXiv:0904.2577 [gr-qc]} {arXiv:0904.2577 [gr-qc]}
  \BibitemShut {NoStop}%
\end{thebibliography}%
\end{document}